\documentclass[10pt]{article}
\usepackage{amsmath,amssymb}
\usepackage[dvipdfm]{graphicx}
\usepackage[english]{babel}
\usepackage{cite,color}
\usepackage[labelfont=bf,labelsep=period,justification=raggedright]{caption}
\makeatletter
\renewcommand{\@biblabel}[1]{\quad#1.}
\makeatother
\date{}
\newtheorem{theorem}{Theorem}

\begin{document}
\begin{flushleft}
{\Large
\textbf{Degree correlations in directed scale-free networks}
}
\\
Oliver Williams$^{1}$, 
Charo I. Del Genio$^{2,3,4,\ast}$
\\
\bf{1} Department of Physics, University of Warwick, Coventry, UK
\\
\bf{2} Warwick Mathematics Institute, University of Warwick, Coventry, UK
\\
\bf{3} Centre for Complexity Science, University of Warwick, Coventry, UK
\\
\bf{4} Warwick Infectious Disease Epidemiology Research (WIDER) Centre, University of Warwick, Coventry, UK
\\
$\ast$ E-mail: C.I.del-Genio@warwick.ac.uk
\end{flushleft}

\section*{Abstract}
Scale-free networks, in which the distribution of the degrees
obeys a power-law, are ubiquitous in the study of complex systems.
One basic network property that relates to the structure of the
links found is the degree assortativity, which is a measure of
the correlation between the degrees of the nodes at the end of
the links. Degree correlations are known to affect both the structure
of a network and the dynamics of the processes supported thereon,
including the resilience to damage, the spread of information
and epidemics, and the efficiency of defence mechanisms. Nonetheless,
while many studies focus on undirected scale-free networks, the
interactions in real-world systems often have a directionality.
Here, we investigate the dependence of the degree correlations on the power-law
exponents in directed scale-free networks. To perform our study, we consider
the problem of building directed networks with a prescribed degree distribution,
providing a method for proper generation of power-law-distributed directed
degree sequences. Applying this new method, we perform extensive numerical
simulations, generating ensembles of directed scale-free networks with exponents
between~2 and~3, and measuring ensemble averages of the Pearson correlation
coefficients. Our results show that scale-free networks are on average uncorrelated
across directed links for three of the four possible degree-degree correlations,
namely in-degree to in-degree, in-degree to out-degree, and out-degree to
out-degree. However, they exhibit anticorrelation between the number of outgoing
connections and the number of incoming ones. The findings are consistent
with an entropic origin for the observed disassortativity in biological
and technological networks.

\section*{Introduction}
The use of networks is fundamental to model the structure
and the dynamics of a vast number of systems found throughout
the natural and engineered worlds. Their main
appeal lies in allowing the reduction of a complex system
to a discrete set of elements, the nodes, that interact across
links. Then, one can study the structural properties of a
network and infer results on the behaviour of the system
thus modelled~\cite{Boc06,Boc14}. The simplest global structural
attribute of a network is its degree distribution
$P\left(k\right)$, which expresses the probability of having
a node with $k$ links. A particularly important case is that
of scale-free networks, in which the degree distribution
obeys a power-law $P\left(k\right)\sim k^{-\gamma}$~\cite{Alb02,New03,Cal07,Del11}.
Scale-free networks have been observed in citation distributions~\cite{Pri65,Red98,New01},
Internet and WWW topology~\cite{Alb99,Vaz02}, biological systems~\cite{Jeo00,Jeo01},
technological, economic and social systems~\cite{Ama00,Lil01},
and transport processes~\cite{Tor04,Tor08},
and therefore they have been the subject of a considerable
body of research.
A generalization of the simple network model
can be introduced by defining a directionality for the links.
Directed networks are more suited to represent systems
in which the interaction between elements is not necessarily
symmetric, such as food webs or gene regulatory networks~\cite{Moe13}.
In this case, the connectivity of a node is no longer represented
by a single scalar, as each node has a number of incoming
connections (its in-degree $k^-$) and a number of outgoing
connections (its out-degree $k^+$).
A related quantity is the degree assortativity, often called
simply assortativity, which measures the tendency
of a node to be connected to nodes of similar degree. Assortativity
is effectively a measure of the correlations amongst node
degrees. As such, it is known to have substantial effects on the dynamical
processes taking place on a network. For instance, assortative
networks are more resistant to fragmentation in case of attack,
while disassortative networks are less prone to cascading failures~\cite{New02,Mas02,New03_2}.
Degree correlations also play an important role in mathematical
epidemiology, as they directly affect the dynamics of epidemic
spreading, as well as the efficiency of defence mechanisms~\cite{Eub04,Dag12,Del13,You13}.
Numerous studies have shown that social networks are typically
assortative, while biological and technological networks are
disassortative, with links preferentially between nodes of high
and low degree.
In the case of directed networks, one can actually consider
four different degree assortativities across links, as one
can model the dependence of either in-degree or out-degree
of a node on either in-degree or out-degree of its neighbours~\cite{Fos10,Squ11,van13}.
Here, we study how a scale-free structure affects
assortativity in directed networks. In particular, we show
that directed scale-free networks exhibit no in--in, out--out
and in--out correlations, but are anticorrelated in the out--in
assortativity.

\section*{Methods}
To study the preferred correlation structure induced
by scale-freeness in directed networks, we performed
extensive numerics, generating statistical ensembles
of networks with power-law distributed in-degrees and
out-degrees. The generation of directed networks
with given degree distributions involves two distinct
phases. First, extract two sequences of integer numbers
that follow the distributions, and assign these to
the nodes as directed half-links, or ``stubs''. Taken
in pairs, these numbers form a so-called bi-degree sequence,
and correspond to the in-degree and the out-degree of
each node. Then, sample the bi-degree sequence
creating network realizations without self-edges or
multiple edges. A suitable method to perform
this second step is the algorithm discussed in Ref.~\cite{Kim12},
which allows an efficient uniform sampling of the realizations
of a bi-degree sequence. However, not every sequence
of integer pairs can be realized by a simple directed
graph. Thus, before being able to apply the sampling
algorithm, we need to develop a procedure to properly
create bi-degree sequences that admit realizations,
which are referred to as graphical. To do
so, we start from the Fulkerson theorem~\cite{Ful60},
which states the necessary and sufficient graphicality
conditions for bi-degree sequences:
\begin{theorem}\label{fulk}
 A sequence of non-negative integer pairs
$\mathcal{D}=\left\lbrace\left(k_1^-,k_1^+\right), \left(k_2^-,k_2^+\right), \dotsc, \left(k_N^-,k_N^+\right)\right\rbrace$
with $k_1^-\geqslant k_2^-\geqslant\dotsb\geqslant k_N^-$ is graphical if and only if
\begin{gather}
 k_1^- \leqslant N-1,\quad \max_{1\leqslant i\leqslant N}k_i^+\leqslant N-1\label{simple}\:,\\
 \sum_{i=1}^N k_i^- = \sum_{i=1}^N k_i^+\label{handsh}\:,\\
 \sum_{i=1}^x k_i^- \leqslant \sum_{i=1}^x \min\left\lbrace x-1,k_i^+\right\rbrace + \sum_{i=x+1}^N \min\left\lbrace x, k_i^+\right\rbrace\:.\label{interesting}
\end{gather}
\end{theorem}
Theorem~\ref{fulk} can be used to efficiently verify
the graphicality of an extracted sequence using the
particularly fast implementation described in~\cite{Kim12}.
The theorem can be intuitively understood by looking
at the three conditions. Condition~\ref{simple} simply
ensures that no node has a number of incoming or outgoing
stubs that exceeds the number of other nodes. This is
clearly a necessary condition, since each node can connect
to or receive connections from at most all the remaining
nodes. To understand Condition~\ref{interesting}, notice
that the left-hand side is just the number of incoming
stubs in the set consisting of the first $x$ nodes.
Then, consider how to maximize this number. To start
with, take each of the first $x$ nodes and connect them
to all the others in the set. However, note that each
node $i$ can only have as many outgoing connections
as its out-degree $k_i^+$. Thus, if the out-degree of
node $i$ is large enough, it can be connected to all
the remaining $x-1$ nodes; otherwise, it can only be
connected to $k_i^+$ amongst the remaining $x-1$. The
first term in the sum on the right-hand side accounts
for these connections. The second term in the sum has
the same meaning. However, the sum is now taken over
the nodes that are \emph{not} within the first $x$.
Thus, each can connect to at most $x$, rather than $x-1$,
other nodes. Finally, Condition~\ref{handsh} mandates
the total number of incoming stubs equal the
total number of outgoing stubs. This introduces
an important constraint in the generation of graphical
bi-degree sequences. To see why, write the number of
incoming and outgoing stubs as
\begin{equation}
 \sum_{i=1}^N k_i^{\pm} = N\left\langle k^{\pm}\right\rangle\:.
\end{equation}
In general, the power-law exponents
for in-degrees and out-degrees in directed scale-free
networks can be different~\cite{Boc06,Boc14,Alb02,New03,Cal07}. Then, if the out-degrees
scale as ${k^+}^{\gamma^+}$ and the in-degrees as
${k^-}^{\gamma^-}$, for $N\gg1$, it is $\left\langle k^{\pm}\right\rangle \approx \frac{\gamma^\pm-1}{\gamma^\pm-2}$. Thus, one cannot
expect the sums of in-degrees and out-degrees to
be equal, if the respective power-law exponents
are different.
In fact, in the region of interest $2<\gamma\leqslant 3$,
the variance of power-law distributions is unbounded, since
$\left\langle {k^\pm}^2\right\rangle \approx \frac{\gamma^\pm-1}{\gamma^\pm-3}$.
Thus, in this range of exponents, one should not expect
Condition~\ref{handsh} to be satisfied even when choosing the same exponent
for in-degrees and out-degrees.

\begin{figure}[!ht]
\begin{center}
\includegraphics[width=0.51\textwidth]{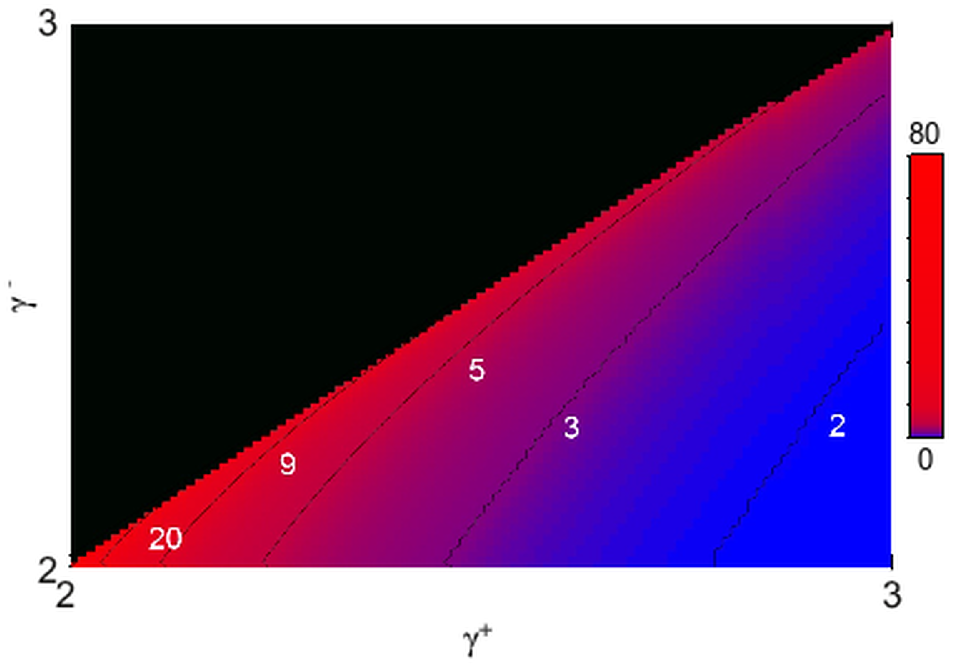}
\end{center}
\caption{
{\bf Effective upper cutoff on the in-degrees if they are conditioned
on the sum of the out-degrees} The contour plot shows the logarithm of
the introduced upper cutoff. Note that for almost all the choices of
power-law exponents, such cutoff is so low that the greatest part of
the distribution tail is lost, affecting the scale-free character of
the resulting network. The labels indicate the logarithm of the cutoff
for the corresponding contour lines. Only half of the region is plotted,
as we are under the assumption that $\gamma^+>\gamma^-$.
}
\label{Fig1}
\end{figure}

To guarantee that Condition~\ref{handsh} is satisfied,
and avoid the trivial non-graphicality of the generated
bi-degree sequence, one cannot extract independently the
sequences of in-degrees and out-degrees. Rather, one should
be extracted without further constraints, and the other
should be conditioned to have the same sum as the former.
Without loss of generality, assume that $\gamma^+>\gamma^-$.
Then, it is $\left\langle k^-\right\rangle>\left\langle k^+\right\rangle$.
Thus, freely extracting the out-degrees requires, on average,
to lower the mean in-degree with respect to its unconstrained
value. This effectively introduces an upper
cutoff excluding all the degrees above a certain threshold
$N_U$. For a large network, the normalization constant
of the in-degree distribution with an upper cutoff is
\begin{equation}
Z_{N_U}\equiv\int_1^{N_U} {k^-}^{-\gamma^-}\ \mathrm{dk^-} = \frac{N_U^{1-\gamma^-}\left(N_U^{\gamma^- -1}-1\right)}{\gamma^- -1}\:.
\end{equation}
Thus, the mean in-degree is
\begin{equation}\label{meanin}
 \left\langle k^-\right\rangle_{N_U} = \int_1^{N_U} \frac{{k^-}^{-\gamma^-+1}}{Z_{N_U}}\ \mathrm{dk^+} = \frac{\gamma^- -1}{\gamma^- -2} N_U \frac{N_U^{\gamma^- -2}-1}{N_U^{\gamma^- -1}-1}\:.
\end{equation}
Equating Eq.~\ref{meanin} with the expression
for the unconstrained mean out-degree yields
\begin{equation}\label{nu}
 N_U \frac{N_U^{\gamma^- -2}-1}{N_U^{\gamma^- -1}-1} = \frac{\gamma^- -2}{\gamma^- -1} \frac{\gamma^+ -1}{\gamma^+ -2}\:.
\end{equation}
The solution to Eq.~\ref{nu}, plotted in Fig.~\ref{Fig1},
show that for almost all the choices of $\gamma^-$ and $\gamma^+$,
the upper cutoff would eliminate the vast majority of the tail
of the degree distribution. As the defining characteristic of scale-free
networks is a power-law tail, this indicates that the choice of
conditioning the in-degree distribution on the sum of the out-degrees
is not suitable for sequence generation.

\begin{figure}[!ht]
\begin{center}
\includegraphics[width=0.51\textwidth]{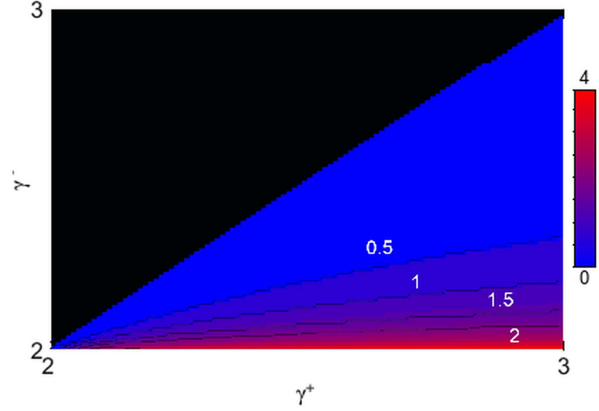}
\end{center}
\caption{
{\bf Effective lower cutoff on the out-degrees if they are conditioned
on the sum of the in-degrees} The contour plot shows the logarithm of
the introduced lower cutoff. Unlike what happens with the reverse choice,
the cutoff introduced is always minor, and it actually vanishes for most
of the choices of in-degree and out-degree exponents. The labels indicate
the logarithm of the cutoff for the corresponding contour lines. Only
half of the region is plotted, as we are under the assumption that $\gamma^+>\gamma^-$.
}
\label{Fig2}
\end{figure}

The other possibility is extracting the in-degrees
in an unconstrained way, and conditioning the out-degrees
on their sum. This time, the cutoff introduced is a lower
cutoff $N_L$ on the out-degree distribution. For $N\gg1$, the
out-degree normalization constant with lower cutoff is
\begin{equation}
Z_{N_L}\equiv\int_{N_L}^\infty {k^+}^{-\gamma^+}\ \mathrm{dk^+} = \frac{N_L^{1-\gamma^+}}{\gamma^+ -1}\:.
\end{equation}
Then, the mean out-degree is
\begin{equation}
 \left\langle k^+\right\rangle_{N_L} = \int_{N_L}^\infty \frac{{k^+}^{-\gamma^++1}}{Z_{N_L}}\ \mathrm{dk^+} = \frac{\gamma^+ -1}{\gamma^+ -2} N_L\:.
\end{equation}
As the two mean degrees have to be equal, it is
\begin{equation}
 N_L = \frac{\gamma^- -1}{\gamma^- -2} \frac{\gamma^+ -2}{\gamma^+ -1}\:.
\end{equation}
Note that, defining the excess exponent $E\equiv \gamma^+ -\gamma^-$,
the equation above can be rewritten as
\begin{equation}
 N_L = \frac{\gamma^- -1}{\gamma^- -2}\left(1-\frac{1}{\gamma^- -1+E}\right)\:.
\end{equation}
This form explicitly shows that $N_L$ is~1 when the exponents are equal and $E=0$,
and it increases monotonically with $E$, towards an asymptotic value
of $\left(\gamma^- -1\right)/\left(\gamma^- -2\right)$. As illustrated
in Fig.~\ref{Fig2}, such cutoff is very mild. Thus, this approach leaves
the tail of the distribution entirely untouched. Moreover, for more than
half of the region of interest, the whole out-degree distribution has no
effective cutoff at all.

Notice that defining a proper method for the generation
of power-law distributed directed degree sequences is essential
for the accuracy of research outcomes. In fact, approximate
techniques have uncontrolled errors and produce results
that depend on the details of the approximation made~\cite{Moe13,Dro09}.

\section*{Results and discussion}
At the light of the considerations expressed in the previous
section, we generated ensembles of bi-degree sequences of random
power-law distributed integers with exponents between~2 and~3,
conditioning the sequence with the greater exponent on the sum
of the sequence with the lower one. Then we tested the sequences
for graphicality, and sampled the graphical ones using the direct
construction algorithm detailed in Ref.~\cite{Kim12}. For each
sample, we measured the assortativities using the Pearson coefficients
\begin{equation}
 r^{\alpha\beta} = \frac{\sum_{i,j}A_{i,j}\Big(k_i^\alpha-\left\langle k^\alpha\right\rangle_L\Big)\left(k_j^\beta-\left\langle k^\beta\right\rangle_L\right)}{\sqrt{\sum_ik_i^\alpha\Big(k_i^\alpha-\left\langle k^\alpha\right\rangle_L\Big)^2}\sqrt{\sum_jk_j^\beta\left(k_j^\beta-\left\langle k^\beta\right\rangle_L\right)^2}}\:,
\end{equation}
where the averages are taken over all directed
links, the elements of the adjacency matrix of
the network $A_{i,j}$ are~1 if there is a link
from node $i$ to node $j$, and $\alpha$ and $\beta$
can be $-$ or $+$, indicating in-degrees or out-degrees,
respectively~\cite{Tel14}. We stress that the
sampling method used is a degree-based graph construction
algorithm~\cite{Del10}. Algorithms
in this class can access the entire space of the
realizations of a graphical sequence. They work
by building the sample graphs via the systematic
placement of links, guaranteeing that the graphicality
of the sequence is maintained after each step.
At every moment, the combinatorially exact probabilities
of placing each allowed link are completely determined.
Thus, these methods efficiently allow
uniform graph sampling, without introducing biases
due to a particular choice of generative model or
construction algorithm, which can result in overrepresentation
or inaccessibility of part of the realization space.

\begin{figure}[!ht]
\begin{center}
\includegraphics[width=0.5\textwidth]{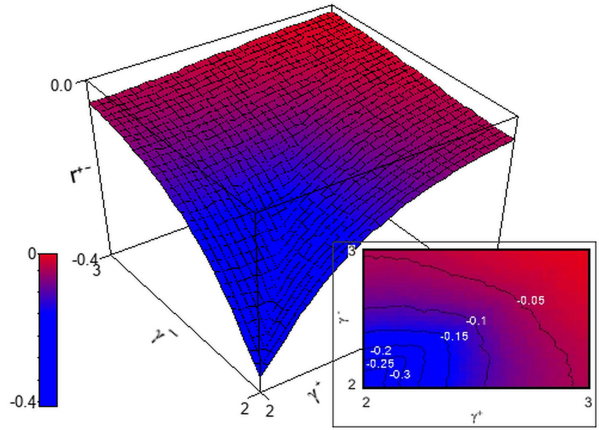}
\end{center}
\caption{
{\bf Degree correlations in directed scale-free networks.}
The Pearson correlation coefficient $r^{+-}$ is always negative,
indicating that directed scale-free networks are naturally
disassortative when one considers the out-in correlation.
The inset shows a contour plot of the same data, for added
clarity. The labels in the contour plot indicate the value
of $r^{+-}$ for the corresponding contour lines.
}
\label{Fig3}
\end{figure}

\begin{figure}[!ht]
\begin{center}
\includegraphics[width=0.55\textwidth]{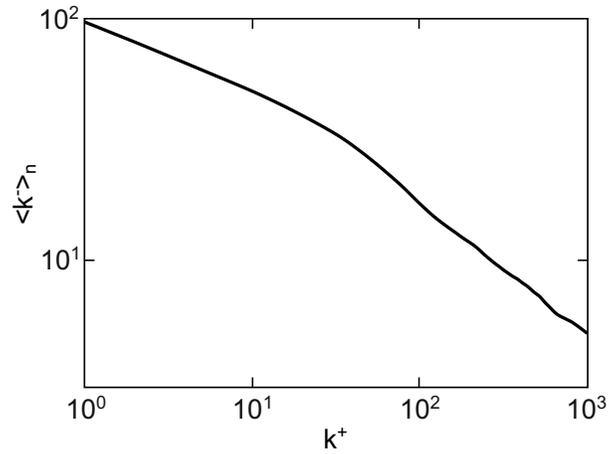}
\end{center}
\caption{
{\bf Disassortative degree correlations.}
The plot shows the average in-degree
$\left\langle k^-\right\rangle_n$ of the
neighbours of nodes with a given out-degree $k^+$
for an ensemble of networks with
$\gamma^+=\gamma^-=2.01$ and $N=1000$.
The dependence of $\left\langle k^-\right\rangle_n$
on $k^+$ clearly indicates that nodes with
low out-degree link preferentially to nodes
of high in-degree, and nodes with high
out-degree link mostly to nodes of low
in-degree. The monotonically decreasing dependence
confirms the strong disassortative nature
of the networks.
}
\label{Fig4}
\end{figure}

\begin{figure}[!ht]
\begin{center}
\includegraphics[width=0.55\textwidth]{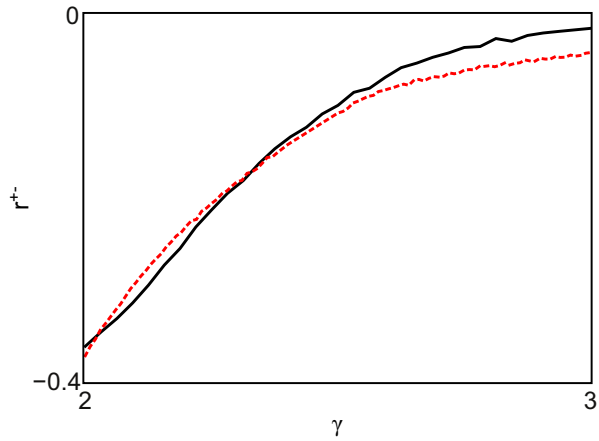}
\end{center}
\caption{
{\bf Entropy-maximizing disassortativity.}
The plot shows the out-in correlation coefficients
for directed scale-free networks with $\gamma^+=\gamma^-\equiv\gamma$.
The simulation data are shown in solid black.
The red dotted line corresponds to the coefficients
that maximize the information entropy for
a given $\gamma$. The good agreement of the
results indicates the entropic origin of the
disassortativity observed in directed scale-free
networks.
}
\label{Fig5}
\end{figure}

The results indicate the absence of any dependence
of the in-in, in-out and out-out coefficients on the
choice of power-law exponents. In fact, these three
coefficients all vanish within the uncertainties throughout
the region studied. Conversely, the out-in coefficient
is always negative (Fig.~\ref{Fig3}), indicating disassortative
correlation between the out-degree of the node at
the beginning of a link and the in-degree of the node
at its end. Figure~\ref{Fig4} illustrates
this pattern of dependence by plotting the average
in-degree $\left\langle k^-\right\rangle_n$ of the
neighbours of nodes with a given out-degree $k^+$,
for an ensemble of networks with $\gamma^+=\gamma^-=2.01$ and $N=1000$. The values
of $\left\langle k^-\right\rangle_n$ decrease quickly and monotonically
with $k^+$, confirming the strong disassortative
nature of the networks. Our results show substantial
similarities between the correlation structure of
directed and undirected scale-free networks. Indeed,
it is a well-known fact that random undirected
scale-free networks are disassortative~\cite{Par03,Mas04,Bog04,Cat05,Lit13}.
Thus, to explain our findings,
we use the entropic treatment described in Ref.~\cite{Joh10},
extending it to directed networks.
To do so, write the information
entropy of a given network as
\begin{equation}\label{entropy}
 S = -{\sum}_{i,j}\left[\varepsilon_{i,j}\log\varepsilon_{i,j}+\left(1-\varepsilon_{i,j}\right)\log\left(1-\varepsilon_{i,j}\right)\right]\:,
\end{equation}
where $\varepsilon_{i,j}$ is the expectation value
for the $\left(i,j\right)$ element of the adjacency
matrix. To derive an expression for $\varepsilon_{i,j}$
in the case of a given bi-degree sequence, note that
it has to satisfy two conditions, namely
\begin{equation}
 \sum_{j=1}^N \varepsilon_{i,j} = k_i^+
\end{equation}
and
\begin{equation}
 \sum_{j=1}^N k_j^-\varepsilon_{i,j} = k_i^+\left\langle k^-\right\rangle_n\left(k_i^+\right)\:.
\end{equation}
A form that satisfies these conditions is
\begin{equation}\label{closed}
 \varepsilon_{i,j} = \frac{1}{N}\left\lbrace\frac{k_i^+ k_j^-}{\left\langle k\right\rangle} + \int\left(\frac{\sigma_2}{\sigma_{\beta+2}}\delta\left(\nu-\left(\beta+1\right)\right)-\delta\left(\nu-1\right)\right)\left[\frac{\left(k_i^+ k_j^-\right)^\nu}{\left\langle {k^-}^\nu\right\rangle} - {k_i^+}^\nu - {k_j^-}^\nu + \left\langle {k^-}^\nu\right\rangle\right]\mathrm{d\nu}\right\rbrace\:,
\end{equation}
where $\beta$ is a free parameter. In principle, the factor
$\left(\frac{\sigma_2}{\sigma_{\beta+2}}\delta\left(\nu-\left(\beta+1\right)\right)-\delta\left(\nu-1\right)\right)$
in Eq.~\ref{closed} can be replaced by any arbitrary function
of $\nu$. However, the choice made, in which
$\sigma_t\equiv\left\langle {k^-}^t\right\rangle - \left\langle k\right\rangle\left\langle {k^-}^{t-1}\right\rangle$,
allows to reproduce the observed dependence of $\left\langle k^-\right\rangle_n$
on $k^+$.
Then, computing the integral in Eq.~\ref{closed}, it is
\begin{equation}\label{epsexp}
 \varepsilon_{i,j} = \frac{1}{N}\left\lbrace\frac{\left\langle {k^-}^2\right\rangle - \left\langle k\right\rangle^2}{\left\langle {k^-}^{\beta+2}\right\rangle - \left\langle k\right\rangle\left\langle {k^-}^{\beta+1}\right\rangle} \left[ \frac{\left(k_i^+ k_j^-\right)^{\beta+1}}{\left\langle {k^-}^{\beta+1}\right\rangle} - {k_i^+}^{\beta+1} - {k_j^-}^{\beta+1} + \left\langle {k^-}^{\beta+1}\right\rangle\right] + k_i^+ + k_j^- - \left\langle k\right\rangle\right\rbrace\:.
\end{equation}
Using Eqs.~\ref{entropy} and~\ref{epsexp}, we can find the choice
of $\beta$ that maximizes $S$, and compute the Pearson
coefficient $r^\ast$ corresponding to the maximum entropy
network for any given power-law exponents. Notice that in the equations
above, we make no distinction between $\left\langle k^+\right\rangle$
and $\left\langle k^-\right\rangle$, as they must be equal to ensure
graphicality of the bi-degree sequence. Also, we restrict the parameter
search to the values that yield networks without self-edges
or multiple edges.
To carry out the calculation, we use the degree-maximizing
sequence as representative of the scale-free
networks for each value of $\gamma$~\cite{Del11}.
Figure~\ref{Fig5} displays a comparison between
$r^{+-}$ as measured by simulations and $r^\ast$,
in the case of $\gamma^+=\gamma^-$.
The two sets of results are substantially in
agreement, save for higher values of $\gamma$,
where $r^\ast<r^{+-}$. This can be explained
by considering that the degree-maximizing sequences
used to compute $r^\ast$ feature more high-degree nodes than would
be found on average, thus decreasing the assortativity
of their realizations.

In summary, we showed that directed scale-free
networks are naturally uncorrelated when considering
in-in, in-out and out-out correlations. Thus, when
looking across a directed link, the in-degree of
the originating node has no influence on the in-degree
or the out-degree of the target node. Similarly,
the out-degrees are not affected by the out-degrees
of the neighbours. However, the out-in correlation
coefficient is found to be negative throughout the
region studied. This indicates that the natural state
of directed scale-free networks is one in which nodes
of low degree prefer to link to nodes of high
degree, and vice versa. The origin of this preference
is entropic, as the coefficients found are in good
agreement with those corresponding to the maximum
information entropy. Thus, the
observation of a disassortative directed scale-free
network is not sufficient to infer the existence of
extra growth mechanisms
beyond those responsible for its degree distribution.
These results suggest that
the disassortative correlations observed in many
real-world systems, such as biological and technological
networks, do not necessarily arise because of design or evolutionary
pressure. In fact, the absence of such drivers,
and the resulting randomness, would lead to the
observation of the anticorrelated state as the
most probable one. Notice that this
does not exclude the presence of evolutionary
mechanisms, which may certainly be the cause of an
observed disassortative network topology in some specific cases.
However, their action would have to promote
the maximum-entropy state, thus making their presence
undetectable from the degree distribution and correlations alone.

\section*{Acknowledgments}
The authors would like to thank Alex Arenas for fruitful discussions.
OW acknowledges support by the URSS (Undergraduate Research
Support Scheme) of the University of Warwick. CIDG acknowledges
support by EINS, Network of Excellence in Internet Science,
via the European Commission's FP7 under Communications Networks,
Content and Technologies, grant No.~288021.

\section*{Figures}

\end{document}